# RECURRENCE ANALYSIS OF VEGETATION TIME SERIES AND PHASE TRANSITIONS IN MEDITERRANEAN RANGELANDS

TEODORO SEMERARO[a], NORBERT MARWAN[b], BRUCE K. JONES[c], ROBERTA ARETANO[a], MARIA RITA PASIMENI[a], IRENE PETROSILLO[a], CHRISTIAN MULDER[d,*] AND GIOVANNI ZURLINI[a,*]

[a] *Landscape Ecology Laboratory, Department of Biological and Environmental Sciences and Technologies, University of Salento, Lecce, Italy*

[b] *Potsdam Institute for Climate Impact Research, Telegraphenberg A 31, 14473 Potsdam, Germany*

[c] *Desert Research Institute, 755 East Flamingo Road, Las Vegas, NV 89119, USA*

[d] *National Institute for Public Health and the Environment, 3720BA Bilthoven, The Netherlands*

* *Corresponding authors:* christian.mulder@rivm.nl, giovanni.zurlini@unisalento.it

**Abstract**

Mediterranean rangelands should be conceived as socio-ecological landscapes (SEL) because of the close interaction and coevolution between socio-economic and natural systems. A significant threat to these Mediterranean rangelands is related to uncontrolled fires that can cause potential damages due to the reduction or even the loss of ecosystems. Our results show that time series of forest and grassland for unburned and burned areas are characterized by both periodic and chaotic components. The fire event caused a clear simplification of vegetation structures as well as of SEL dynamics that is more regular and predictable after the burning and less chaotic. However grassland evolution could be more predictable than forest considering the effect of fire disturbance on successional cycles and stages of the two land-cover types. In particular, we applied recurrence analysis with sliding temporal windows three-year length on the original time series. This analysis indicates that grasslands and forests behaved similarly in correspondence with the burning, although their phase states slowly diverge after fire. Recurrence is useful to study the vegetation recovery as it enables mapping landscape transitions derived from remote sensing. The approach helps stakeholders to draw landscape interventions and improve management strategies to sustain the delivery of ecosystem services.

**Introduction**

Complex adaptive systems (Levin 1998) like Mediterranean rangelands should be conceived as socio-ecological landscape (SELs) (Berkes and Folke 1998; Berkes *et al.* 2003; Zaccarelli *et al.* 2008) because of the close interaction and coevolution between socio-economic and natural systems (Mulder *et al.* 2015; QUINTESSENCE Consortium 2016). SELs show generally a nonstationary and complex behaviour, with their usual phase state fluctuating around some average. They are commonly assumed to respond smoothly to gradual change in climate, habitat fragmentation or exploitation. Such a condition can be sporadically interrupted by an abrupt shift to a radically different regime (Schlesinger *et al.* 1990; Scheffer and Carpenter 2003; Westley *et al.* 2006). Transformations from one state to another state correspond to ecological transitions between alternative phases. When the integral structure of the systems is changed (Li 2000) it results in changes in the kind and level of ecosystem services provided by the SELs.

Our ability to evaluate with accuracy the complexity of natural and human-controlled processes from reliable dynamical measures is therefore crucial from the perspective of ecosystem management and restoration. Thus, especially in regions under drought, the identification of recurrent behaviours or irregular cycles, tipping points and proximity to transition to desertification is critical. In this respect, one of the most critical challenges is to understand how historical dynamic profile of SELs evolved in response to internal processes (e.g., plant succession) and external drivers (e.g., climate change). Those profiles can tell us a great deal about past and current SEL dynamics (Walker *et al.* 2002; Antrop 2005; Zurlini *et al.* 2006) and how the system might respond in the future (Walker *et al.* 2002). To this purpose, satellite remote sensing provides the only means of monitoring dynamic land-cover profiles and fire events at local through regional up to global scales (Goetz *et al.* 2005; Röder *et al.* 2008).

We applied non-linear time series techniques such as Recurrence Quantification Analysis (RQA) (Marwan *et al*. 2007) to the analysis of spatial-temporal dynamics of land-cover classes in Mediterranean rangelands. We used time series of vegetation indices to better understand interannual variability of land-cover dynamics in relation to fire disturbance, recovery and stability/predictability.

Time series analyses of ecological indicators describing non-linear landscape behaviour focused on different aspects of temporal complexity like the identification of changes and the determination of discontinuities using different linear transformation analysis (Jassby and Powell 1990; Rodríguez-Arias and Rodó 2004). Several approaches have been proposed for analysing image time series, such as Principal Component Analysis (Crist and Cicone 1984), wavelet decomposition (Anyamba and Eastman 1996), Fourier analysis (Azzali and Menenti 2000) and Change Vector Analysis (Lambin and Strahler 1994).

Entropy and information theory have been extensively applied in ecology (Ulanowicz 2001), including environmental assessments (Ekström 2003; Magurran 2004), evolution (Avery 2003), and species coexistence (Chen *et al.* 2005; Parrott 2005). Spectral entropy has been used to characterize relative order in time of a variety of land-cover systems based on Fourier transformation of the original signal (Zaccarelli *et al.* 2013; Zurlini *et al.* 2013).

These time series analyses discriminate noise from the signal by its temporal characteristics but involve some type of manipulation of the original time series like transformations designed to isolate dominant components of the variation across years of imagery across the multi-temporal spectral space. However, because of nonstationarity, noise, relatively short time series (Strang 1994), and complex shapes of recurrent cycles, we need enhanced methods for time series analysis, such as quantifying non-linear dynamics (Marwan 2008) to foster predictability and to identify dynamical transitions (Scheffer *et al.* 2009).

One promising method of non-linear time series analysis is RQA, developed for investigating the complex behaviour of dynamical systems and working quite well even with nonstationary short time series (Guimarães-Filho *et al.* 2010; Donges *et al.* 2011; Marwan *et al.* 2015). The strength of the RQA approach resides in its independence from constraining assumptions (outliers, linearity, stationarity) and signal transformations (detrending, noise filtering) (Marwan *et al*. 2007, 2015) and in operating directly on the original series without any data manipulation. The practical and powerful use of RQA in the study of complex, time-varying dynamical systems has been demonstrated by interdisciplinary applications, such as for cardiovascular health diagnosis, behavioural, cognitive and neurological studies, studying fluid dynamics and plasma, analysing optical effects, or palaeoclimate regime change detection (Marwan *et al.* 2007; Marwan 2008). However, RQA has never been applied to landscape stability analysis.

"Recurrence" reveals all the times when the phase space trajectory of the dynamical system visits roughly the same area in the phase space and therefore it recurs. Such cyclic patterns provide useful indications on the resilience capacity of a SEL in a retrospective way exploring the ability of the system to absorb disturbances occurred in the past (Zurlini *et al.* 2006, 2007).

In this study, the comparison of burned and unburned areas through RQA demonstrates the impacts of fire and the recovery capacity of land covers to pre-burn levels, and to indicate changes in functional variability associated with vegetation succession consistent with early successional species. We provide examples of this approach and discuss what resulting RQA implies in terms of phase transition identification and description, and what they tell about the change in behaviour of different land covers.

**Study area**

Rangelands in Apulia (SE Italy) are concentrated in the Gargano peninsula with forested areas, besides few remnants of evergreen forests interspersed with olive groves, and natural drylands like semi-arid grasslands, garigue and steppes and permanent pastures. The study area covers 1500 ha in the municipality of Vieste and it is located inside the Gargano National Park. This area, in particular, is characterized by pinewood (*Pinus halepensis* Mill) mixed with Mediterranean xeric grasslands (Thero-Brachypodietea) (Biondi *et al.* 2010), that were subjected to large fires intentionally started in July 2007. Pine forests are autochthonous and naturally occurring in the area where there is no harvest of trees and a long history of grazing and local fires. The area belongs to a typical Mediterranean semi-arid region (Pueyo and Alados 2007). Unfavourable biophysical factors include erratic precipitation (mainly during the winter), high summer temperature with frequent drought events, poor and erodible soils, extensive deforestation with frequent fires and land abandonment (Ladisa *et al.* 2012). After the fire event, the vegetation grew spontaneously (i.e.

secondary succession) as the study area was not subjected to interventions of environmental restoration.

**Materials and methods**

We have applied non-linear time series techniques such as Recurrence Quantification Analysis (RQA) (Marwan *et al.* 2007) at time series of vegetation indices to analyze the persistence or transition of structures and functions of the forest in relation to fire disturbance and land-cover recovery in terms of stability/predictability. The time series were built using Enhanced Vegetation Index (EVI), extracted by MODIS imagery from 2000 to 2014 given by 323 16-day composite MODIS images (Terra MOD13Q,1) with 250 m resolution (USGS, 2014). So, one data time is equal to 16 days. Pixels were selected and defined as forests or grasslands when the relative cover was higher than 70%.

EVI is an ecological functional surrogate of above-ground net primary production (ANPP) (Xiao *et al.* 2004) that is a key-controlling factor determining most of provisioning ecosystem services. ANPP is a measure of the available solar energy captured by the system and is, therefore, an indicator of the overall ecosystem functioning (Costanza *et al.* 2007). The EVI is calculated as follows:

$$EVI = 2.5 * (NIR - RED) / (NIR + (6 * RED - 7.5 * BLUE) + 1) \qquad (1)$$

where the NIR is the reflectance or radiance in the near infrared channel; RED and Blue are the reflectance or radiance in the visible channel.. The inclusion of the blue band for atmospheric correction is an important feature to study areas where seasonal burning of pasture and forest takes place throughout the dry season, either for agricultural purpose or for natural accidents (Xiao *et al.* 2004). To better understand interannual variability of land-cover types in the study area, we applied RQA to the EVI time series profiles of forest and grassland burned areas as well of surrounding unburned areas with similar vegetation composition as reference. The method is based on the Recurrence Plot (RP) that represents an advanced technique of nonlinear data analysis. It is a visualisation of a square matrix, in which the matrix elements correspond to those times at which a state of a dynamical system recurs (columns and rows correspond then to a certain pair of times). Technically, the RP reveals all the times when the phase space trajectory $\vec{x}_i$ of the dynamical system visits roughly the same area in the phase space.

The idea behind this recurrence analysis is that recurrence is a fundamental property in many natural processes and that different dynamics exhibits different but characteristic recurrence properties. In order to visualize such recurrences, Eckmann et al. (1987) have introduced the RP. The recurrences are investigated on the state vectors $\vec{x}_i$ which can be constructed by a time delay embedding (if only one measurement $u_i$ is available, e.g., the EVI) by $\vec{x}_i = \left(u_i, u_{i+\tau}, \ldots, u_{i+(m-1)\tau}\right)$ with $m$ as the final phase space dimension and $\tau$ an appropriate time delay (Packard *et al.* 1980). The time delay $\tau$ should be selected in such a way to minimize autocorrelations between points of the time series, e.g., by using mutual information. The choice of $m$ is usually based on counting false nearest neighbours when increasing $m$, and choosing such value of $m$ where the number of false nearest neighbours vanishes (Marwan *et al.* 2007, 2010). The RP is then based on the recurrence matrix:

$$R_{i,j} = \begin{cases} 1 & \text{if } \|\vec{x}_i - \vec{x}_j\| \leq \varepsilon \\ 0 & \text{otherwise} \end{cases} \quad \text{with } \vec{x}_i \in \mathbb{R}^m \text{ and } i,j = 1,\ldots,N \quad (2)$$

where $N$ is the number of considered states $\vec{x}_i$; $\varepsilon$ is a threshold and $\|\cdot\|$ a norm. Note that $\varepsilon$ is essential, as systems often do not recur exactly to a formerly visited state but just approximately.

The RP is a graphical representation of this recurrence matrix, where black points represent those time points where the spatial distance between two states $\vec{x_i}$ and $\vec{x_j}$ is falling below the threshold ε and therefore the system recurs (Eckmann *et al.* 1987; Marwan *et al.* 2007, 2015). Diagonal lines in the RP indicate that the evolution of states is similar at different times and, thus, can point to deterministic processes; vertical lines in the RP indicate that some states do not change or change very slowly for some time and can suggest laminar or persistent states (Fig. 1).

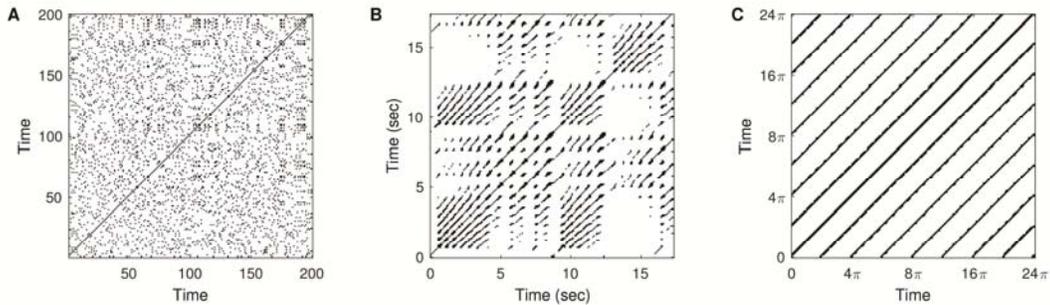

**Fig. 1.** Recurrence plots of typical dynamics: (A) stochastic process (white, uniformly distributed noise), (B) chaotic process (Lorenz system), and (C) periodic process (sine oscillation). In each example 200 time points have been used: (A) no embedding, (B) the 3 components of the Lorenz system, (C) embedding in 2 dimensions with a π/2 lag.

The RQA is using line structures within the RP to derive several measures of complexity like (Marwan *et al.* 2007):

➢ Recurrence rate (RR): the percentage of recurrence points in the RP;

➢ Determinism (DET): the percentage of recurrence points which form diagonal lines in the RP (can be interpreted in terms of predictability);

➢ Laminarity (LAM): the percentage of recurrence points which form vertical lines in the RP (can be interpreted in terms of rate of change of the system). High values of LAM are an indication of dynamics that is trapped more often to certain states;

➢ Divergence (DIV): the inverse of the longest diagonal line. It is an indicator of the divergence rate (chaoticity) of the dynamics. The higher the DIV, the more chaotic (less stable) is the signal.

Abrupt changes in the dynamics as well as extreme events cause white areas or bands in the RP thus allowing finding and assessing extreme and rare events easily by using the frequency of their recurrences. RPs were derived from the mean EVI profiles of burned forests and grasslands, while RQA was devised in three steps. In the first step, RR, DET, LAM and DIV were calculated for each single pixel of forests and grasslands belonging to the burned area considering the time series from 2000 to 2014. Each measure is then spatially averaged, hence providing one mean value of RR, DET, LAM and DIV. Comparisons between mean values were conducted through simple T-tests.

Embedding dimension and time delay were derived from the analysis on different single pixels of forest and grassland and choosing their more representative values that were for embedding dimension = 3, and time delay = 1. As to the recurrence threshold, we have used a fixed threshold with the maximum norm. However, for studying dynamical transitions, the threshold selection is not of fundamental importance, because the relative change of RQA measures does not substantially depend on it (Marwan *et al.* 2015).

In the second step, the RQA analysis was carried out splitting the time series for each pixel considered before and after the fire event. In this step, time delay and embedding dimension were both set to 1 because time series were too short. Additionally, the RQA can provide useful insights in the different dynamical behaviour between burned and unburned areas. For this purpose, in the third and last step the joint recurrence analysis (JRA) (Marwan *et al.* 2007) was applied using pixels of similar land covers from burned and surrounding unburned areas. JRA has the advantage, that the data can be different observables and can have different variability and variation. It can be used for the detection of phase synchronization or general synchronization.

The joint recurrence plot (JRP) shows that two (or more) systems recur simultaneously to the neighbourhood of a formerly visited point in their respective phase space. Formally, the joint recurrence matrix is the result of the element-wise product of two (or more) recurrence matrices of the data series (Romano et al., 2004; Marwan *et al.* 2007), showing all the times at which a recurrence in one dynamical system occurs simultaneously with a recurrence in the second dynamical system:

$$JR_{i,j} = R_{i,j}^{\text{forests}} \cdot R_{i,j}^{\text{grassland}} \qquad (3)$$

where $R_{i,j}^{\text{forests}}$ and $R_{i,j}^{\text{grassland}}$ are the entries at (*i*,*j*) in the recurrence matrices of the forests system and grassland system. The JRP is finally the graphical representation of $JR_{i,j}$ where black dots represent the coincidence of recurrences between the forests and grassland systems.

Next, the JRP was performed in a way that allows investigating the temporal change of the recurrence properties. In order to do this, we applied sliding temporal windows of two and three year length on the original time series. This windowing procedure comes with the drawback of reducing the number of observations. Therefore, time delay and embedding dimension were both set to 1. RP, DET and LAM were also calculated from the JRP. All the analyses were carried out in MATLAB using the CRP Toolbox available at http://tocsy.pik-potsdam.de/CRPtoolbox/.

**Results**

Overall, EVI time series of forest and grassland for unburned and burned areas are characterized by both periodic and chaotic components (Figs. 2 and 3). For the burned area, the phase transition owing to the fire event and the recovery processes are quite evident. After the fire, the periodic component prevails for both forests and grasslands likely because of the binding of primary productivity of post-fire land cover to seasonal climatic variations, but the frequency of variation is lower (see the discussion section).

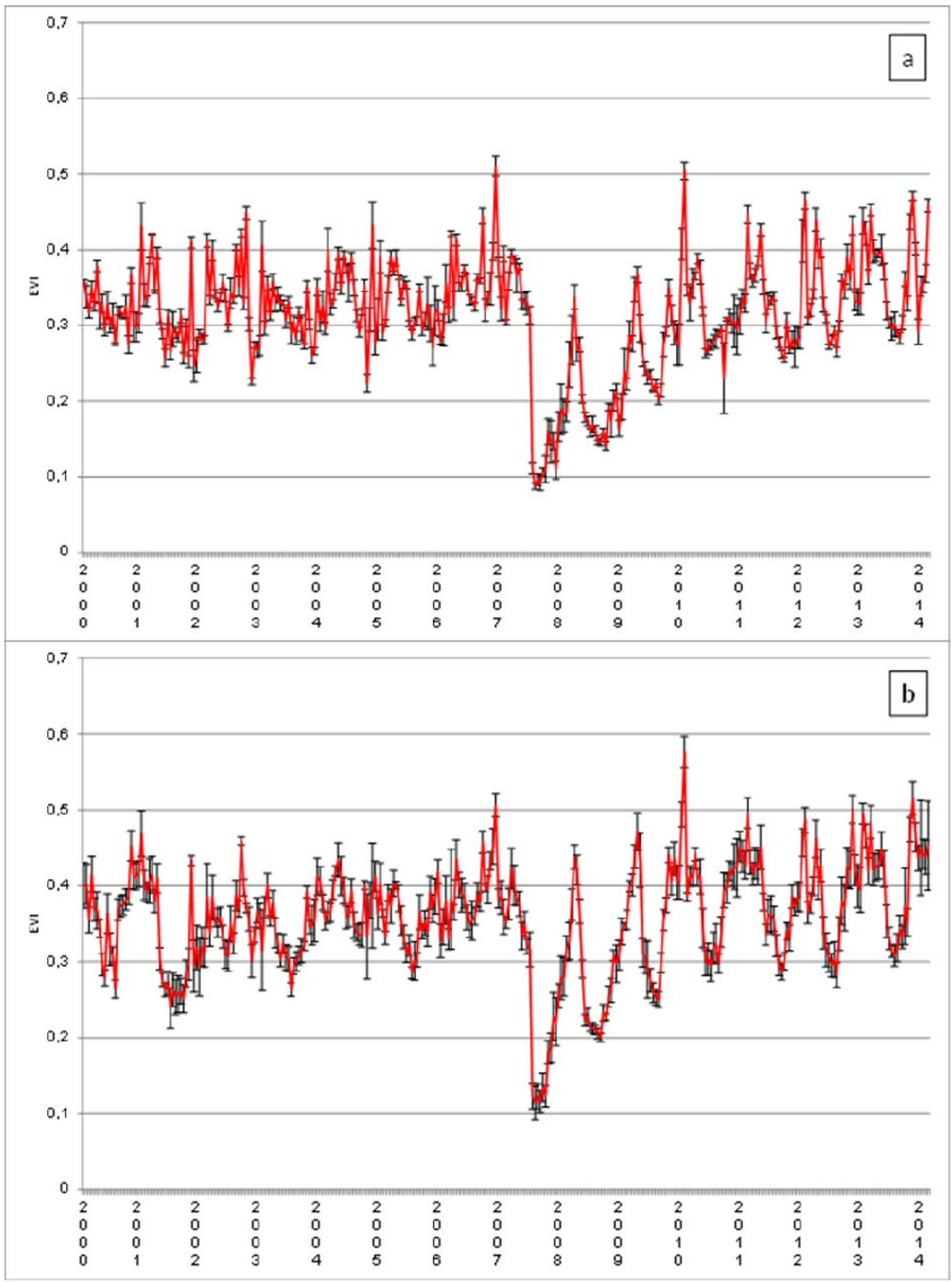

**Fig. 2.** Mean time series of the Enhanced Vegetation Index (EVI) for burned forests (a) and grasslands (b) from 2000 to 2014 in the Gargano National Park rangelands showing the fire event and the recovery processes. Standard error bars are shown.

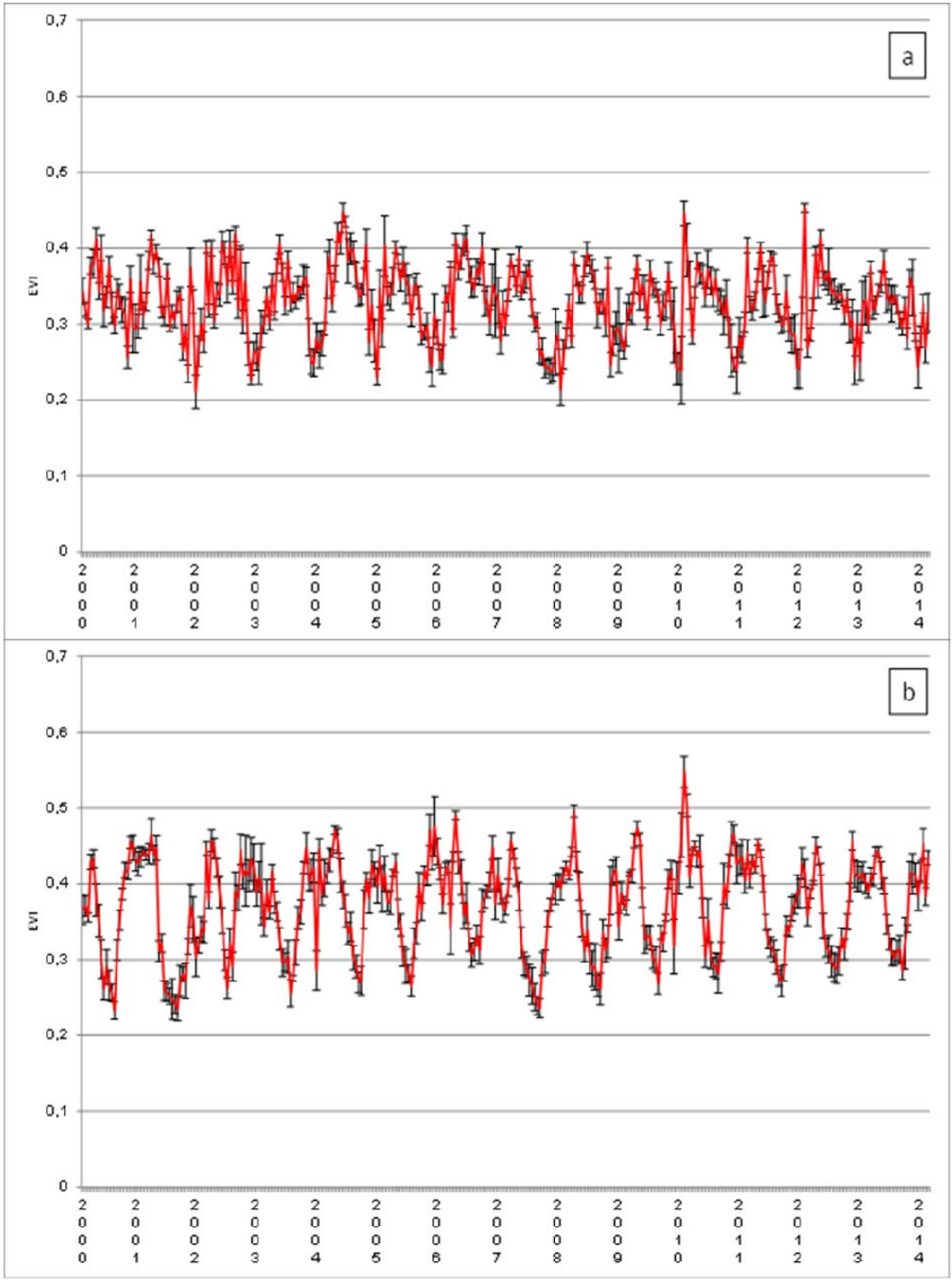

**Fig. 3.** Mean time series of the Enhanced Vegetation Index (EVI) for unburned forests (a) and grasslands (b) from 2000 to 2014 in the Gargano National Park rangelands. Standard error bars are shown.

Correspondingly, patterns in the RPs for forest and grassland burned areas reveal phase transitions between July 2007 and January 2009 (Fig. 4) identified by "disruptions" (white bands) in all RPs in correspondence with the fire event of July 2007. After that, systems tend to return to the original stability domain. However, for both land use we visually find periodic patterns in the corresponding RPs, revealing mainly the seasonal variability, but for the grasslands we find a more erratic or chaotic pattern than for forests.

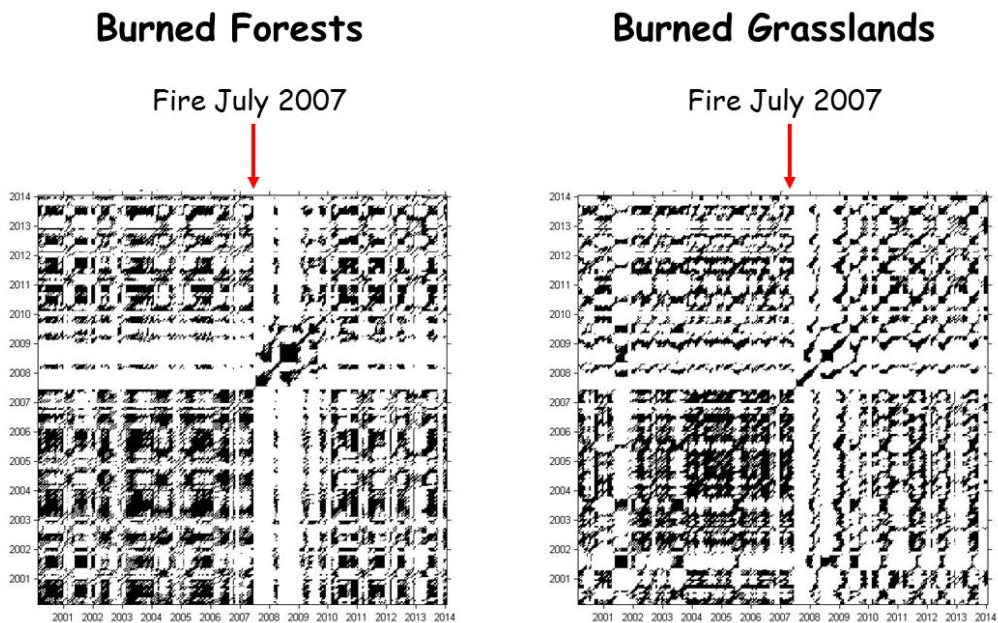

**Fig. 4.** Recurrence plot (years vs. years) of forest and grassland satellite image time series for the burned areas in the Gargano National Park. Phase transitions between July 2007 and January 2009 are identified by "disruptions" (white bands).

Next, we calculated the RQA measures for these RPs over the entire period (2000 to 2014). Moreover, to better analyse the pattern following the phase transition owing to the fire event, the EVI time series of forests and grasslands were split in two sub-series: before (February 2000 to July 2007) and after (January 2009 to February 2014) the fire event excluding the transition phase. In both cases, for each pixel of the satellite image we calculated RQA measures and averaged these values over the area. A summary of the relevant recurrence measures for forests and grasslands in burned areas are given in Table 1 for the overall time series, and in Table 2 for the two sub-series.

**Table 1.** Mean Recurrence Rate (RR), Determinism (DET), Laminarity (LAM) and Divergence (DIV) for the overall time series for burned land covers.

| Parameters | Forests | | Grasslands | | |
|---|---|---|---|---|---|
| | Mean | SD | Mean | SD | **T-test** |
| **RR** | **0.2977** | 0.0001 | **0.2972** | 0.0002 | **P > 0.05** |
| **DET** | **0.913** | 0.006 | **0.926** | 0.008 | **P < 0.05** |
| **LAM** | **0.818** | 0.019 | **0.864** | 0.016 | **P < 0.05** |
| **DIV** | **0.0299** | 0.005 | **0.029** | 0.004 | **P >0.05** |

**Table 2.** Mean Recurrence Rate (RR), Determinism (DET), Laminarity (LAM) and Divergence (DIV). The analysis was carried out splitting the original time series in two: before and after the fire event excluding the transition phase.

| Parameters | Before Fire | | | | | After Fire | | | | |
|---|---|---|---|---|---|---|---|---|---|---|
| | Forests | | Grasslands | | | Forests | | Grasslands | | |
| | Average | SD | Average | SD | T-Test | Average | SD | Average | SD | T-Test |
| RR | **0.2954** | 0.0001 | **0.2954** | 0.0001 | P>0.05 | **0.2947** | 0.0002 | **0.2948** | 0.0001 | P>0.05 |
| DET | **0.515** | 0.010 | **0.600** | 0.037 | P<0.05 | **0.720** | 0.028 | **0.721** | 0.045 | P>0.05 |
| LAM | **0.599** | 0.046 | **0.719** | 0.038 | P<0.05 | **0.823** | 0.022 | **0.824** | 0.036 | P>0.05 |
| DIV | **0.110** | 0.014 | **0.102** | 0.014 | P>0.05 | **0.072** | 0.073 | **0.082** | 0.021 | P>0.05 |

For the overall time series, RR and DIV were not significantly different between forests and grasslands, whereas DET and LAM show significant differences. In particular, DET for grasslands is higher indicating more predictable dynamics than such for forests. Also LAM is higher for grasslands, indicating that grassland states change slower than forest states. From these results we conclude that grassland evolution could be more predictable than forest considering the effect of fire disturbance on successional cycles and stages of the two land-cover types.

Interestingly, after the fire event, means for RR, DET, LAM and DIV for forests and grasslands are not significantly different (Table 2), likely because of large grass understorey in early successional forests. There is also a marked increase in DET and LAM for forests and grasslands and a slight decrease in DIV when compared to the before-burning situation. Some of these metrics support what is shown in Figs. 2 and 3, namely that there is a longer frequency between variability intervals after the fire. Thus, forests and grasslands present a rather similar behaviour after the burning event with a clear transition from less regular/predictable to more regular/predictable dynamics/evolution.

Transitions from forests to grasslands and from the old to the new states of grasslands caused by the fire event can be highlighted using the JRA to compare the evolutionary behaviour of EVI profiles for burned and unburned forest and grassland areas with sliding windows of two and three years. By applying JRA it is possible to identify joint disruptions in correspondence of the fire event (Fig. 5a). Joint-LAM and Joint-DET for burned areas reveal transitions from less to higher determinism and laminar states in correspondence with fire followed by a slow decrease to less deterministic and laminar states. Joint-DET (Fig. 5b,d) and Joint-LAM (Fig. 5c,e) evolve similarly after the burning.

We applied JRA also to grassland and forest time series from neighbouring unburned reference areas. In this case Joint RQA shows no evidence of disruption/transition (Fig. 6a). Also the profiles of Joint-DET (Fig. 6b,d) and Joint-LAM (Fig. 6c,e) do not show relevant variation indicating that each land cover maintained its proper functional traits.

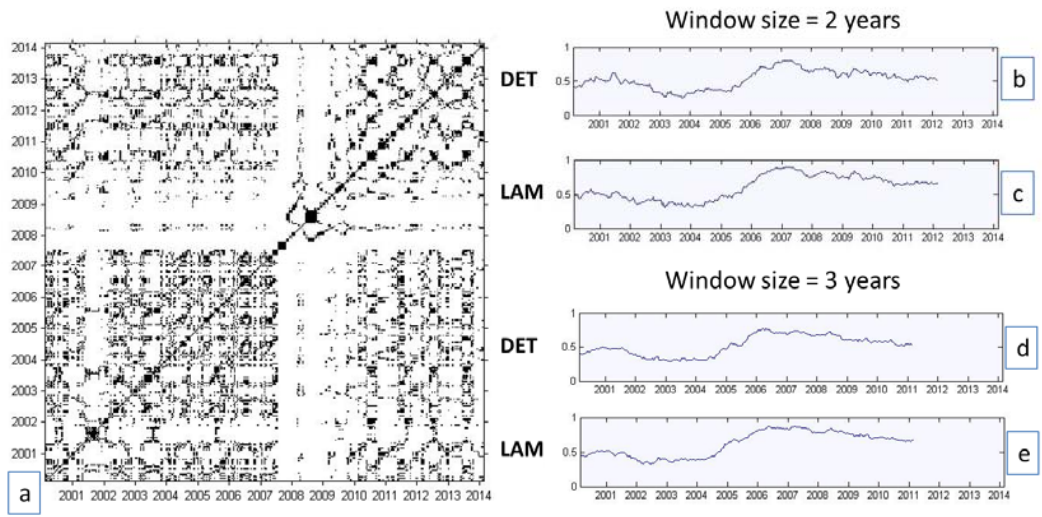

**Fig. 5.** Joint Recurrence Analysis of burned forests and grassland time series. Joint recurrence plot (a); time series for two and three years moving window size for Joint-DET (b and d) and Joint-LAM (c and e). Joint phase transitions are identified by white bands (a).

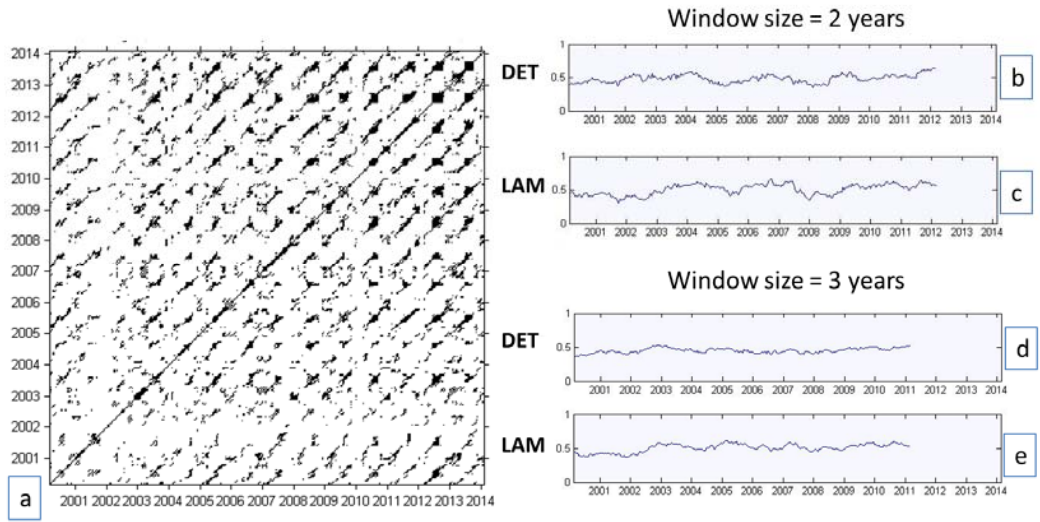

**Fig. 6.** Joint Recurrence Analysis of unburned forest and grassland time series. Joint recurrence plot (a); time series for two and three years moving window size for Joint-DET (b and d) and Joint-LAM (c and e).

## Discussion

In Mediterranean drylands, vegetation recovery following fire events always starts with grass and herbs, which often recolonize quickly as a result of the higher nutrient availability and their immediate ability to cover open spaces (e.g., Christensen 1994). With sprouting or seeding being the main regeneration strategies based on the "ecological memory" of the system, plant communities in Mediterranean rangelands commonly develop through shrub-dominated populations (gariga) with single, isolated trees at a later stage, to eventually dense matorrals or evergreen forests, given sufficiently long-term undisturbed conditions (Fig. 7).

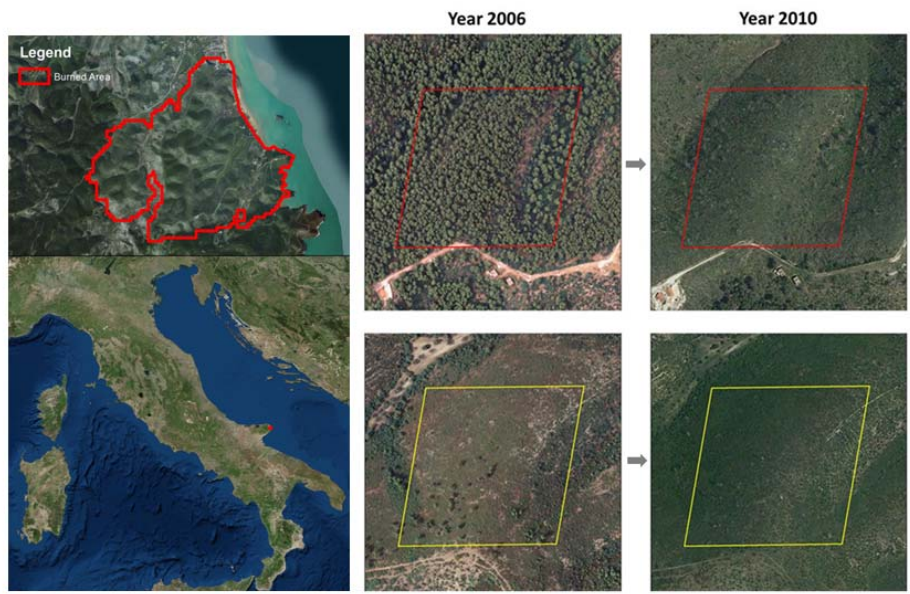

**Fig. 7.** Phase transitions in Gargano National Park rangelands (left panel). Upper right, forested area (2006) to grassland (2010); lower right, between grasslands. The MODIS pixel extent is shown inside the burned area of ortophotos with a 50 cm resolution as online at http://www.sit.puglia.it/portal/portale_cartografie_tecniche_tematiche/WMS.

The manifestation of post-fire secondary succession is determined by a combination of local physical and climatological conditions, spatial variability in burn severity, the vegetation cover composition, and the presence or absence of exogenous disturbance factors during the phase of recovery (e.g., Quinn 1994; Whelan 1995; Lloret and Vilà 2003; Eidenshink *et al.* 2007; Rollins 2009). The frequency of fires in Mediterranean rangelands has historically promoted the adaptation of evergreen trees like *Pinus halepensis* and grasses like

*Hyparrhenia hirta* and *Brachypodium ramosum* whose fire resilience depends on adaptive protective mechanisms as well as life-history and recovery traits (e.g. Noble and Slatyer 1980; Keeley 1986; Quinn 1994).

Characterization of fire severity might further improve our understanding of potential trajectories of grassland and forests, although caution should be applied in how these will be defined (Keeley 2009). Moreover, creating expected-to-observed dynamic state and phase transitions from biophysical data (soil, elevation, climate, etc.) could enhance the detection of state-space changes (Wylie *et al.* 2012). Transitions observed in RPs are due to the destruction of the original vegetation cover during the burning in both forest and grassland areas. The fire event caused a clear simplification of vegetation structures as well as of SEL dynamics that is more regular and predictable after the burning (i.e., higher DET) and less chaotic (lower DIV).

While forest vegetation is replaced by new grassland vegetation with coniferous seedlings, grassland areas preserve the same grassland structure after the burning with similar type of xeric vegetation composition because of the resistance to fire of the structuring herbaceous species. Because of the frequency of local fires in such rangelands to counter the onset of shrubs and encourage grazing, no phase transition of forest to scrubland or grassland to scrubland occurred that could greatly reduce primary productivity and the delivery of ecosystem services by the landscape.

The RQA approach can form an efficient strategy for visualizing and quantifying temporal dynamics in SELs, and to gauge phase transition like those observed from forests to grasslands but providing more accurate and novel quantitative insights. RQA and JRA show that both transformation and persistence in grasslands and forests work together allowing living systems to quickly recover and return to the previous specific functional states.

Even if the time interval of the analysis is too short for an exhaustive knowledge of the future evolution of vegetation covers after the burning, results of JRA can provide a useful insight into their natural evolutionary behaviour. High values of Joint-DET and Joint-LAM in correspondence with the burning indicate that grasslands and forests behaved quite similarly. However, after that, a slow decrease in time for Joint-DET and Joint-LAM of grasslands and forests (Table 2) indicates that the probability that both systems recur simultaneously in the same way is lowering. That means that the evolution of the phase states of the two land covers slowly diverge after fire before returning to their original states. Efforts to reduce the risk of undesirable state transitions to foster predictability in SELs should address the gradual changes that affect resilience or each property linked with resilience, rather than control perturbations (Scheffer and Carpenter 2003; Müller *et al.* 2016). RQA is potentially capable of describing both gradual and abrupt transitions in land cover and provide an effective alternative way to study various aspects linked to resilience in real SELs. This is an explorative approach to understand what type of new insight and information recurrence analysis could provide to help the landscape planner to find good strategies or actions in landscape planning. The advantage of recurrence analysis compared to other methods is that it can be implemented without transforming the original series that often implies to exclude information considered as noise. Then the recurrence analysis allows analysing in synergic way the different components that make up the time series, such as cyclicality, seasonality, trends and unpredictable components that simultaneously can be altered by a perturbation, determining what will be the overall system behaviour. The potential practical utility of this approach mainly resides in the mapping of transitions in the landscape based on time series of state variables like EVI, which could help planners and managers formulate specific landscape management strategies in a spatially explicit manner. Thus, the mapping of recurrence variables most sensitive to transitions like DET and LAM based on EVI time series would help to gauge spatially explicit predictability gradients of transitions in the landscape. This could be of great support to planners and managers to draw specific landscape management strategies for sustaining ecosystem services (e.g., Butterfield and Malmstrom 2006; Jones *et al.* 2013). From such predictability maps of invariant landscape functions, more suitable structural/functional connection network and fragmentation could be designed for more effective conservation strategies. On the other hand, the mapping of low predictable areas because of transitions in vegetation indices could visualize change concentrations in SELs indicating to planners where the predictability should be increased through a proper change in management or type of vegetation cover in light of both impeding further land degradation and fostering the overall conservation network.

**Acknowledgments**

Piero Medagli and two referees are gratefully acknowledged for their suggestions; the authors also acknowledge the ERASMUS funding program and one DFG grant to N. Marwan (RTG 2043/1, Natural Hazards and Risks in a Changing World).